\documentclass{Interspeech2024}

\usepackage{pgfplots}
\usetikzlibrary{backgrounds}
\usepackage{subcaption}
\usepackage{multirow}
\usepackage{multicol}
\usepackage{tabularx}

\interspeechcameraready

\title{Enhancing CTC-based speech recognition with diverse modeling units}
\name{Shiyi Han, Zhihong Lei, Mingbin Xu, Xingyu Na, Zhen Huang}

\address{
  Apple Inc.}
\email{\{shiyi\_han, zlei, mingbinxu, zhen\_huang, xingyu\_na\}@apple.com}

\begin{document}

\maketitle
 
\begin{abstract}
In recent years, the evolution of end-to-end (E2E) automatic speech recognition (ASR) models has been remarkable, largely due to advances in deep learning architectures like transformer. 
On top of E2E systems, researchers have achieved substantial accuracy improvement by rescoring E2E model's N-best hypotheses with a phoneme-based model. This raises an interesting question about where the improvements come from other than the system combination effect.
We examine the underlying mechanisms driving these gains and propose an efficient joint training approach, where E2E models are trained jointly with diverse modeling units. This methodology does not only align the strengths of both phoneme and grapheme-based models but also reveals that using these diverse modeling units in a synergistic way can significantly enhance model accuracy. Our findings offer new insights into the optimal integration of heterogeneous modeling units in the development of more robust and accurate ASR systems.
\end{abstract}{Automatic Speech Recognition (ASR), Connectionist
Temporal Classification (CTC), Joint training}

\section{Introduction}

In recent years, both academia and industry have increasingly adopted end-to-end (E2E) ASR models \cite{xu2023conformer, xu2023training}, which predict graphemic units such as wordpieces given audio. One of the production-friendly structures \cite{DBLP:conf/interspeech/YaoWWZYYPCXL21, DBLP:conf/interspeech/ZhangWPSY00YP022} of the E2E system is a multi-pass system that uses a Connectionist Temporal Classification (CTC) \cite{graves2006connectionist} model in the first pass and an Attention-Encoder-Decoder (AED) \cite{DBLP:conf/emnlp/ChoMGBBSB14, DBLP:journals/corr/BahdanauCB14} in the second pass.
While E2E ASR does not require a pronunciation lexicon, researchers have successfully enhanced the accuracy of end-to-end speech recognition with pronunciations. One approach is to incorporate pronunciations into the modeling units of the E2E system, such as \cite{xu2019improving,li2022phonetic,higuchi2022hierarchical, DBLP:conf/interspeech/MaHYHH22}. Without having to change the modeling units, another approach maps rare named entities to common wordpieces through pronunciation to boost the recognition of these named entities, such as \cite{lei2024personalization, DBLP:conf/interspeech/HuangALE20}.

In addition to these works, a standalone phoneme-based acoustic model (AM) can be leveraged, for example, AM fusion \cite{DBLP:conf/asru/LeiXHLHNZPHDS23}. The idea of AM fusion is to incorporate an external AM into the E2E system in a similar way to language model (LM) shallow fusion\cite{kannan2018analysis}.  
The authors adopt a strategy by interpolating scores across multiple recognition passes. In these passes, the N-best hypotheses are generated from the wordpiece E2E model, which are then re-ranked in the rescoring pass. 

Despite considerable improvement from AM fusion, it is unclear yet whether AM fusion can generalize to other modeling units. Also, one can argue that the improvement is primarily from the system combination effect and additional model parameters in the separate AM.

To answer the first question, we started by generalizing AM fusion from phoneme to character. We confirm that a character-based AM achieves similar improvement over a wordpiece-based E2E model. AM fusion with either phoneme or character achieved a relative word error rate (WER) reduction of 7.9\% each on test-clean, and 5.6\%, 6.0\% respectively on test-other.

As an answer to the second question, we attempted to remove the need of a separate model by encoder sharing. More specifically, we jointly trained a wordpiece-phoneme/character model by adding an extra CTC output on an intermediate encoder layer in the baseline wordpiece model. This leads to a training criterion with two CTC losses and a wordpiece AED loss which are linearly combined. We then ran first pass wordpiece CTC decoding. Our experiments show that even without phoneme/character rescoring, the first pass wordpiece CTC WER was improved by up to 7.1\% and 6.4\% on test-clean and test-other, respectively. This joint training approach does not just provide a simple first pass alternative to AM fusion with similar improvement, but also confirms that the improvement from AM fusion can be achieved by simply providing representations of smaller units without an increase in model size or runtime computation.

To study the aforementioned improvement in details, we conducted an empirical study to understand the information each encoder layer learns. The result is in line with the findings of \cite{DBLP:conf/iclr/ShimCS22}, which utilized attention visualization to demonstrate that initial layers of a model tend to learn pronunciation-related information, while subsequent layers focus on textual information. We also proves that intermediate CTC loss helps produce robust models which is aligned with \cite{9414594} but our method uses different units and yields superior results. To generalize our finding on not only alphabetic language, we also experimented with logographic language and find similar trend as above.

Our main contributions are as follows: 1. We propose a new joint training strategy that leverages the naturally learned intermediate representations of the network.
2. We investigate the optimal layer matching between different units, resonating with previous studies on the interpretability of speech models.
3. We extend the AM fusion work by improving the Word Error Rate (WER) with models based on alternative units, such as characters.
\section{Methodology}

\subsection{Backbone Model}

Our backbone model follows the recipe of \cite{DBLP:conf/interspeech/YaoWWZYYPCXL21, DBLP:conf/interspeech/ZhangWPSY00YP022, DBLP:conf/interspeech/GulatiQCPZYHWZW20}. Consider $T$ represents frame number and $r$ is the subsampling factor, a conformer based encoder is used to map the audio feature $\mathbf{X} \in \mathbb{R}^{m \times T_1}$ to high-level representation $\mathbf{H}^{L+1} \in \mathbb{R}^{n \times T_2}$, where $T_2 = \lfloor T_1 / r \rfloor$.
To combat the large vocabulary in real-life application, people tend to derive a wordpeice based sequence $Y$ to make the output space tractable.
The encoder is jointly trained by minimizing the CTC objective $f_{CTC}$ and the AED objective $f_{AED}$, and they are parameterized by $\theta_{CTC}$ and $\theta_{AED}$ respectively:
\begin{multline}
    \mathbf{Loss}(X, Y) = f_{CTC}(H^{L+1}, Y; \theta_{CTC}) + \\
                \lambda_{AED} \cdot f_{AED}(H^{L+1}, Y; \theta_{AED}).
    \label{eq:ctc-aed}
\end{multline}
During inference, CTC prefix beam search \cite{graves2006connectionist} is used to find the most reasonable hypothesis $\hat{Y}$ within the beam constraint and AED scoring is usually employed to further improve the accuracy.

\subsection{Acoustic Model Fusion}
On top of the above, AM-fusion firstly trains a frame-synchronous phoneme classifier with an arbitrary structure like conformer or transformer.
We use $H'$ and $L'$ to emphasize that it's trained separately with its own parameters. 
In inference, the top hypothesis is chosen by the weighted summation of the first pass score (CTC in this backbone model) and the score from phoneme-based model:
\begin{multline}
    TopHyp(X) = \arg \max_{\hat{Y}}
        \big[ f_{CTC}(H^{L+1}, \hat{Y};\theta_{CTC}) + \\
         \lambda_{ph} \cdot f_{ph}(H'^{L'+1}, \hat{Y}_{ph}; \theta_{ph}) \big].
    \label{eq:amf}
\end{multline}
where $\lambda_{wp}$ is determined by some held-out data.
The primary objective of AM fusion lies in integrating an external phoneme-based AM into an E2E ASR system. Notably, \cite{DBLP:conf/asru/LeiXHLHNZPHDS23} has demonstrated a considerable enhancement in accuracy through this approach with a standalone AM fusion model.

While both wordpiece and phoneme-based AMs operate on a frame-synchronous basis, aligning the word boundary introduces challenges, making its implementation non-trivial. To overcome this, the authors adopt a different strategy by interpolating scores across multiple recognition passes. In these passes, the top-K hypotheses are generated using either CTC prefix beam search or Weighted Finite State Transducer \cite{mohri2002weighted} (WFST) search. Subsequently, these hypotheses are re-ranked in the rescoring pass using Equation \ref{eq:amf}. 

In this paper, we proposed to further explore the possibility to use different modeling units such as phoneme, character and logographic decomposition when applicable. 

\subsection{Multiple modeling units joint training}
One evident drawback associated with multi-pass recognition is the irrecoverability of early sub-optimal decisions. 
In particular, the challenge arises in the recognition of rare words, where the statistical prominence of wordpieces learned during training tends to overshadow the uncommonly spelled/pronounced words.
To tackle this challenge, we propose a approach involves training the model with an additional CTC output layer followed by an objective focused on other modeling units like characters and phonemes. 
Given a set of decomposition of interest, we augment the overall objective function by appending the CTC loss corresponding to each representation which leads to Equation \ref{eq:tr_wp_ph_i}:
\begin{multline}
    Loss(X, Y) = f_{CTC}(H^{L}, Y; \theta_{CTC}) \ +\\
        \lambda_{AED} \cdot f_{AED}(H^{L}, Y; \theta_{AED}) \ +\\
        \sum_{unit} \lambda_{unit} \cdot f_{CTC}(H^{l_{unit}}, Y_{unit}; \theta_{CTC_{unit}}).
    \label{eq:tr_wp_ph_i}
\end{multline}
In the described equation, the term ${unit}$ denotes the specific unit chosen to enhance the base model, incorporating alternative transcription representations such as phonemes, characters, etc. It is also worth noting that the backbone CTC and AED are trained to condition on the last encoder layer, while the $CTC_{unit}$ is trained on an intermediate layer, where $0 < l <= L$. The notation $H^{l_{unit}}$ represents using different layers of encoder output as the input to the selected unit's $CTC_{unit}$ layer. By using different layers of encoder output for different $CTC_{unit}$, we can achieve the joint training purpose. During decoding, $CTC_{unit}$ is not involved. Both $CTC_{unit}$ and the AED can be used in the rescoring stage or they can both be individually used in the first pass decoding. 
The benefits of our proposed approach are manifold. 
Firstly, leveraging external knowledge acquired during training enables its direct application in the initial pass, without introducing additional inference time complexity.
Secondly, the synergy between the semantic knowledge embedded in wordpieces and the phonetic knowledge inherent in phonemes is seamlessly learned.
Thirdly, the incorporation of multitask training streamlines the application of external knowledge, eliminating the necessity for multi-pass recognition. 
This does not only enhance the efficiency of the system but also reduces latency.
Finally, the elimination of the standalone phoneme-based model confirms that the improvement from AM fusion comes from the pronunciation information rather than combination of two distinct systems. As a bonus, the space for an extra AM is saved for a deeper and likely more robust model, further contributing to overall system improvement.

\begin{figure}[]
\centerline{\includegraphics[width=0.5\linewidth]{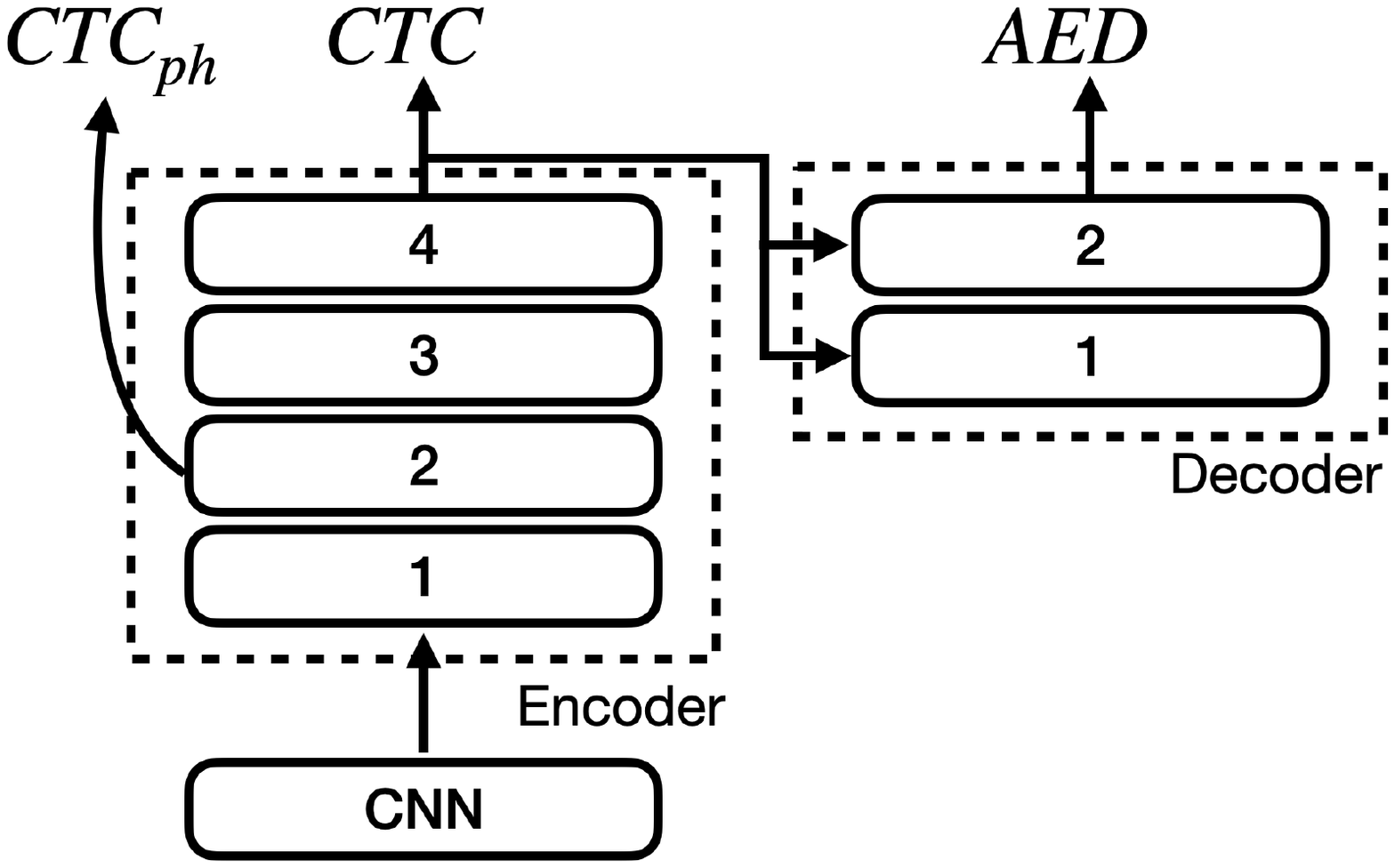}}
\caption{An illustration of the proposed model with 4 encoder layers and 2 decoder layers. A phoneme $CTC_{ph}$ is added on top of the 2nd encoder layer.}
\label{model}
\end{figure}
While Equation \ref{eq:tr_wp_ph_i} outlines a plausible approach that incorporates secondary knowledge, our experiment indicates that the resulting improvement was less than that achieved by applying the standalone model when the extra unit is trained on the top encoder layer. 
Although counter-intuitive, this observation aligns with findings from \cite{DBLP:conf/iclr/ShimCS22}.
In their study, two metrics, cumulative attention diagonality (CAD) and phoneme attention relationship (PAR), are introduced to analyze and quantify the linguistic and phonetic modeling capabilities of each layer. 
Both metrics converge to the same conclusion: lower layers excel at extracting phonetic features, while upper layers are adept at aggregating linguistic information.
Furthermore, adopting our notation, \cite{DBLP:conf/iclr/ShimCS22} constructs a series of classifiers, denoted as $[C_0, C_1, ... C_{L-1}]$, utilizing outputs from different hidden layers $[H^1, H^2, ..., H^{L}]$ respectively. 
The accuracy curve concerning the layer index $l$ exhibits a bell-shaped pattern, with $H^{(L + 1)/2}$ ($L = 17$) yielding the most precise phoneme classification.

Such observations suggest that the output of the last hidden layer may not necessarily be the optimal choice for serving the additional objective. Instead, the multitask objective should be extended to encompass the appropriate representation of audio features, denoted as $H^{l}$.
Furthermore, it is crucial that the objective function aligns the right representation of the audio with the correct representation of the transcripts to maximize its effectiveness. 
This necessitates a careful pairing of the audio and text representations within the objective function.

It is worth noting that there exist alternative methods for representing the transcripts, utilizing a different set of symbols beyond phonemes and wordpieces.
We categorize them into 3 groups:
\begin{description}
    \item[Orthographic] The orthographies of words are decomposed into smaller subword units, such as characters or wordpieces. Byte Pair Encoding (BPE) \cite{DBLP:conf/acl/SennrichHB16a} stands out as one of the most widely used representatives to obtain wordpieces. 
    \item[Phonetic] The phonetic decomposition is derived by linguists, though in practice, often obtained from a grapheme-to-phoneme (G2P) model. The pronunciation dictionary in hybrid ASR \cite{mohri2002weighted}, the character spelling of phonograms, and Pinyin (a romanization system representing pronunciation) in Chinese fall under this category.
    \item[Logographic] This form of decomposition finds applicability primarily in logographic languages, where the written system often reflects semantic composition. An illustrative example is the Wubi input method, which decomposes Chinese characters based on their structural components and maps each character to a unique sequence of keystrokes.
\end{description}

Moreover, our observations indicate that certain units (${unit}$) require mapping to an appropriately matched encoder layer to optimize outcomes, a finding that echoes the insights of the aforementioned paper \cite{DBLP:conf/iclr/ShimCS22}. This alignment between specific units and encoder layers underscores the nuanced relationship between model architecture and the efficacy of joint training, highlighting the strategic importance of layer selection in achieving our stated aims.

\begin{figure*}[h]
\begin{subfigure}[b]{0.42\linewidth}
\centering
\resizebox{\linewidth}{!}{
    \begin{tikzpicture}[inner frame sep=0]
    \begin{axis}[
        width=8cm, height=4.5cm,
        axis lines=left,
        xlabel={layer index $l$},
        x label style={at={(axis description cs:0.5,0.03)}},
        ylabel={WER},
        y label style={at={(axis description cs:0.06,.5)}},
        xmin=0, xmax=13,
        ymin=3, ymax=3.9,
        xtick={0,1,2,3,4,5,6,7,8,9,10,11,12},
        legend style={at={(0.55,1.0)},anchor=north,legend columns=2},
        ymajorgrids=false,
        grid style=dotted,
    ]

    \addplot[color=cyan, mark=none, domain=0:13, densely dashed] {3.67}; \label{curve:baseline:CTC}
    \addplot[color=red, mark=square] table[x=idx, y=clean-ctc]{results-wp-ch.dat}; \label{curve:char:CTC}
    \addplot[color=blue, mark=triangle] table[x=idx, y=clean-ctc, skip coords between index={10}{13}]{results-wp-ph.dat}; \label{curve:ph:CTC}

    \addplot[color=magenta, mark=none, domain=0:13, loosely dashed] {3.24}; \label{curve:baseline:CTC-AED}
    \addplot[color=red, mark=square*] table[x=idx, y=clean-a-rescore]{results-wp-ch.dat};\label{curve:char:CTC-AED}
    \addplot[color=blue, mark=triangle*] table[x=idx, y=clean-a-rescore]{results-wp-ph.dat}; \label{curve:ph:CTC-AED}

    
    \end{axis}
    \end{tikzpicture}
}
\caption{test-clean}
\label{fig:wp_ch_clean}
\end{subfigure}
\begin{subfigure}[b]{0.6\linewidth}
\centering
\resizebox{\linewidth}{!}{
    \begin{tikzpicture}[inner frame sep=0]
    \begin{axis}[
        width=8cm, height=4.5cm,
        axis lines=left,
        xlabel={layer index $l$},
        x label style={at={(axis description cs:0.5,0.03)}},
        ylabel={WER},
        y label style={at={(axis description cs:0.06,.5)}},
        xmin=0, xmax=13,
        ymin=8, ymax=9.9,
        xtick={0,1,2,3,4,5,6,7,8,9,10,11,12},
        legend cell align={left},
        legend style={at={(1.0, 0.5)},anchor=west,legend columns=1},
        ymajorgrids=false,
        grid style=dotted,
    ]

    \addplot[color=cyan, mark=none, domain=0:13, densely dashed] {8.76}; 
    \addplot[color=red, mark=square] table[x=idx, y=other-ctc]{results-wp-ch.dat}; 
    \addplot[color=blue, mark=triangle] table[x=idx, y=other-ctc, skip coords between index={10}{13}]{results-wp-ph.dat}; 

    \addplot[color=magenta, mark=none, domain=0:13, loosely dashed] {8.53};
    \addplot[color=red, mark=square*] table[x=idx, y=other-a-rescore]{results-wp-ch.dat};
    \addplot[color=blue, mark=triangle*] table[x=idx, y=other-a-rescore]{results-wp-ph.dat};

    \legend{\small $CTC$, \small {\ \ \ \ + $CTC_{char}^{j}$}, \small {\ \ \ \ + $CTC_{ph}^{j}$}, \small $CTC$ \& $AED^{j}$, {\ \ \ \ + $CTC_{char}^{j}$}, \small {\ \ \ \ + $CTC_{ph}^{j}$}}
    \end{axis}
    \end{tikzpicture}
}
\caption{test-other}
\label{fig:wp_ch_other}
\end{subfigure}
\caption{WER on LibriSpeech when different $CTC_{unit}$s are attached.}
\label{fig:lbs}
\end{figure*}
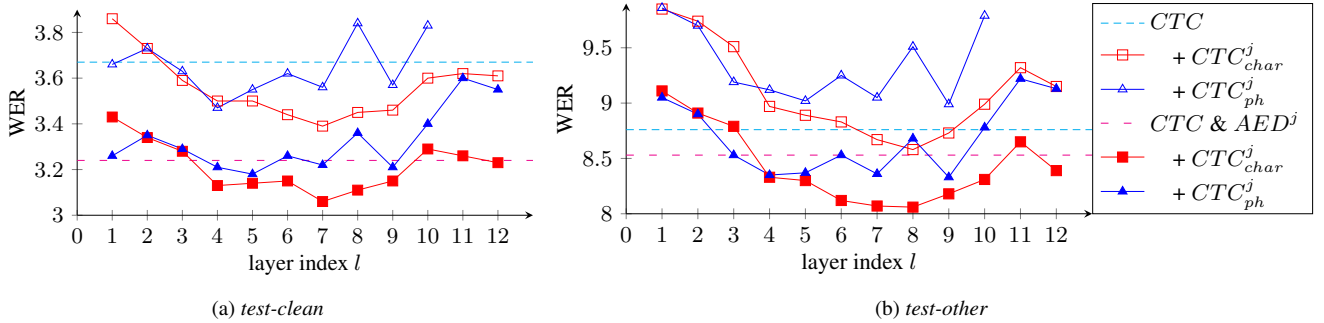

\section{Experiments}

\subsection{Overview}
In this work, we use a conformer CTC-AED model as our baseline modeling structure. This model employs a 12-layer conformer encoder and a 6-layer transformer decoder, each layer using a 4-head self-attention with 256 units. In addition, the decoder layers attend to the encoder outputs using a 4-head, 256-unit cross-attention. All models feature a 4x subsampling convolution block which consumes 80-dimensional log-Mel filterbank features obtained from a 25ms window with a 10ms stride. This model has a total of 47M parameters.

We conduct experiments on English and Chinese Mandarin. We use LibriSpeech \cite{panayotov2015librispeech} to train our English models, while AISHELL-2 \cite{du2018aishell} for Chinese Mandarin. All baseline English models use a vocabulary of 5000 wordpieces, constructed with the SentencePiece library \cite{DBLP:conf/emnlp/KudoR18}, meanwhile Chinese transcripts are tokenized into characters.

\subsection{LibriSpeech}
We start by training a baseline wordpiece CTC-AED model using the aforementioned architecture, a phoneme CTC (without AED) model and a character CTC (without AED) model. The phoneme/character model use the same encoder architecture as the baseline model. The phoneme representation is sourced from \cite{DBLP:conf/interspeech/LugoschRITB19, DBLP:conf/interspeech/McAuliffeSM0S17}, encompassing 70 distinct symbols.
While the character representation is inherently self-contained, we follow the SentencePiece library, treating initial characters and character within a word as distinct units, even though they are originally the same character.
The quality of the system is measured by word error rate (WER).
For the rest of the evaluation, if the rescoring model is not explicitly stated, a wordpiece CTC prefix beam search is assumed. 
It is worth noting that the phoneme CTC model is evaluated in conjunction with an n-gram LM provided by LibriSpeech to produce transcription.

We then follow \cite{DBLP:conf/asru/LeiXHLHNZPHDS23} to rescore the N-best hypotheses from the wordpiece model using either the phoneme or the character CTC. Results in Table \ref{tab:lbs} show a significant relative WER reduction, 7.9\% (3.65 $\rightarrow$ 3.36) each on test-clean, and 5.6\% (9.26 $\rightarrow$ 8.76), 6.0\% (9.26 $\rightarrow$ 8.70) on test-other, using phoneme/character CTC respectively, despite the fact that the phoneme/character CTC is much weaker than the wordpiece baseline.

\begin{table}[h]
    \centering
    \begin{tabular}{l|l|r|r}
        \hline
        \multirow{2}{*}{Loss} & \multirow{2}{*}{Rescoring model} & \multicolumn{2}{c}{WER} \\
        \cline{3-4}
        & & clean & other \\
        \hline
        \multicolumn{4}{c}{Separately trained AM}  \\
        \hline
        $CTC_{char}$ &  - & 3.93 & 10.53 \\
        $CTC_{ph}$ & - & 4.46 & 10.26 \\
        \hline
        \multirow{4}{*}{$CTC$ + $AED^{j}$} & - & 3.65 & 9.26 \\
        & $AED^{j}$ & 3.21 & 8.45 \\
        & $CTC_{mono}^{s}$ & 3.36 & 8.74 \\
        & $CTC_{char}^{s}$& 3.36 & 8.70 \\
        \hline
        \multicolumn{4}{c}{Jointly-trained AM} \\
        \hline
        \multirow{2}{*}{\ \ \ \ \ \ \ \ + $CTC_{ph}^{j}$} & - & 3.53 & 8.76 \\
        & $AED^{j}$  & 3.13 & 8.00 \\
        \hline
        \multirow{3}{*}{\ \ \ \ \ \ \ \ + $CTC_{char}^{j}$} & - & 3.39 & 8.67  \\
        & $AED^{j}$ & 3.06 & 8.07  \\
        & $CTC_{char}^{j}$ & 3.31 & 8.57 \\
        \hline
    \end{tabular}
    \caption{We use superscripts $s$ and $j$ to refer to models trained separately as a standalone and models trained jointly along with the backbone, respectively. When the phoneme model is used in first pass, we combine the lexicon and LM from LibriSpeech in order to properly search.}
    \label{tab:lbs}
\end{table}

Next, we jointly train wordpiece-phoneme models with an additional phoneme CTC output affixed to the 6th layer with an interpolation weight of 0.1 applied to the loss. Without introducing extra modeling parameters or rescoring pass, the model achieved 3.3\% (3.65 $\rightarrow$ 3.53) WER reduction on test-clean, and 5.4\% (9.26 $\rightarrow$ 8.76) WER reduction on test-other. 

We further exhaustively affix the CTC loss of phoneme to various layers $l$. As $l$ increases from encoder layer 1, the WER shows gradual improvement, reaching its peak around the 4th layer before declining. When the loss is attached to the last conformer encoder layer, it starts to hurt word error rate. This trend aligns with the findings of \cite{DBLP:conf/iclr/ShimCS22}, wherein the examination of phoneme modeling capability across different layers similarly demonstrates optimal accuracy in the middle layer. 
With such compelling evidence, we conclude that the middle layer serves as the most appropriate phonetic representation of the deep Conformer model and, consequently, is the optimal location for integrating pronunciation knowledge.

We replicate the previous experiment with a character-CTC.
We observed a more significant improvement, 7.1\% (3.65 $\rightarrow$ 3.39) on test-clean and 6.4\% (9.26 $\rightarrow$ 8.67) on test-other. Following a similar trend, where the peak WER reduction was most pronounced in the 7th layer, which is 3 layers deeper than phoneme's optimal layer. Furthermore, we notice that even if we use last layer of encoder for character-based CTC, we still get an on-par accuracy compared to the base model. Our rationale is that, English, as a alphabetic system, incorporates numerous clues to pronunciation within its written form, allowing characters to carry phonetic information. In comparison to phonemes, characters also contain linguistic information to some extent, as certain combinations of spellings convey specific meanings. When we concatenate the character CTC layer into the middle layers, it helps the model fit the pronunciation representations, thereby improving the word error rate. Additionally, as another representation of textual information, characters hold similar information to wordpieces. When concatenated with the last encoder layer, the accuracy is almost on par.
Nevertheless, phonemes primarily serve as a phonetic representation without specific meanings. We also try to cooperate both phoneme and character loss. It turns out that there is no further improvement, as the secondary knowledge they bring duplicates. 

\subsection{AISHELL-2}

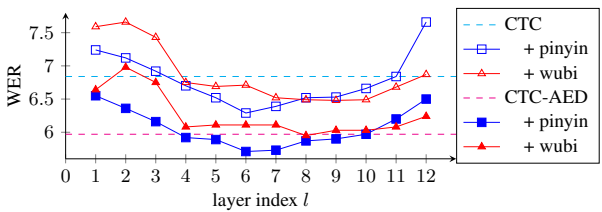
\begin{figure}[h]
\centering
\resizebox{\linewidth}{!}{
    \begin{tikzpicture}[inner frame sep=0]
    \begin{axis}[
        width=8cm, height=4cm,
        axis lines=left,
        xlabel={layer index $l$},
        x label style={at={(axis description cs:0.5,0.03)}},
        ylabel={WER},
        y label style={at={(axis description cs:0.06,.5)}},
        xmin=0, xmax=13,
        ymin=5.6, ymax=7.8,
        xtick={0,1,2,3,4,5,6,7,8,9,10,11,12},
        legend cell align={left},
        legend style={at={(1.0,0.5)},anchor=west,legend columns=1},
        ymajorgrids=false,
        grid style=dotted,
    ]


    \addplot[color=cyan, mark=none, domain=0:13, dashed] {6.84};

    \addplot[color=blue, mark=square] table[x=idx, y=ios-ctc]{aishell2-pinyin.dat}; \label{curve:pinyin:aishell2-ctc}
    \addplot[color=red, mark=triangle] table[x=idx, y=ios-ctc]{aishell2-wubi.dat}; \label{curve:wubi:aishell2-ctc}
    
    \addplot[color=magenta, mark=none, domain=0:13, dashed] {5.97};

    \addplot[color=blue, mark=square*] table[x=idx, y=ios-ctc-aed]{aishell2-pinyin.dat};\label{curve:pinyin:aishell2-ctc-aed}
    \addplot[color=red, mark=triangle*] table[x=idx, y=ios-ctc-aed]{aishell2-wubi.dat}; \label{curve:wubi:aishell2-ctc-aed}


    \legend{\small CTC, {\ \ \ \ + pinyin}, {\ \ \ \ + wubi}, CTC-AED, {\ \ \ \ + pinyin}, {\ \ \ \ + wubi}}
    
    \end{axis}
    \end{tikzpicture}
}
\caption{CER trend on test-ios partition of AISHELL2}
\label{fig:aishell2}
\end{figure}

\begin{table}[]
    \centering
    \begin{tabular}{l|rrr|rrr}
         \hline
         \multirow{2}{*}{Loss} & \multicolumn{3}{c|}{Pure CTC} & \multicolumn{3}{c}{Rescored by AED} \\
         \cline{2-7}
         & ios & and & mic & ios & and & mic \\
         \hline
         Baseline & 6.84 & 7.39 & 7.47 & 5.97 & 6.52 & 6.55 \\
         \ \ \ \ + $CTC_{pinyin}$ & 6.29 & 6.93 & 7.07 & 5.71 & 6.28 & 6.43 \\
         \ \ \ \ + $CTC_{wubi}$ & 6.48 & 7.00 & 7.10 & 5.95 & 6.51 & 6.57 \\
         \hline
    \end{tabular}
    \caption{CER on ios / android / mic partitions of AISHELL2}
    \label{tab:aishell2}
\end{table}

To evaluate the effectiveness of logographic representation in Mandarin, we conduct experiments on the AISHELL-2 dataset \cite{du2018aishell} using Pinyin \cite{duanmu2007phonology} and Wubi, which respectively represent the phonetic decomposition and structural decomposition of Chinese characters. We use a similar model setting as English experiments', a 12-layered conformer with CTC-AED loss. Character error rate (CER) serves as the accuracy metric in the corresponding experiments. Results in Table \ref{tab:aishell2} reveals CER improvements ranging from 5.4\% to 8.0\% using Pinyin without further AED rescoring, higher than improvements from Wubi keystrokes. This is likely because Pinyin introduces more information that is otherwise less well represented by Chinese characters. Notably, when AED rescoring is used, Pinyin still preserves the majority of the gains, while Wubi's effect is marginal. We conduct an exhaustive search for suitable intermediate representations. 
As illustrated in Figure \ref{fig:aishell2}, we find that layer 5 provides the most improvement when jointing training with Pinyin CTC, while layer 9 is the most suitable for Wubi CTC joint training. 
This observation aligns with the premise that the effectiveness of phonetic modeling reaches its peak in the middle layers, whereas the effectiveness of linguistic modeling tends to increase with depth.
Wubi's effectiveness starts to decline after reaching its peak at layer 9, possibly because the word structure does not always fully embed word semantics, and as the model goes deeper, the surrounding context plays a more significant role in representing semantics \cite{harris1954distributional, mikolov2013distributed}.

\subsection{Highlights}

These experiments validate the improvements from additional pronunciation information without any form of system combination. 
More importantly, our method offers additional advantages. It eliminates the need for additional training-time compute resources by avoiding the inclusion of an additional AM.
This reduction in computational demand leads to a decrease in the overall system footprint.
Our approach obviates the necessity for a second pass rescoring to benefit from the accuracy improvements achieved through AM fusion.
Consequently, this reduces both engineering efforts and inference compute resources required during deployment.

\section{Conclusion}

In this work, we achieve training-time knowledge injection by attaching objective functions corresponding to various modeling units. 
We show that these objectives should be linked to the hidden layers that excel at representing the information carried by such units in order to properly inject the knowledge.

\bibliographystyle{IEEEtran}
\bibliography{mybib}

\end{document}